\documentclass{ws-procs9x6}

\begin{document}
\begin{center}
Accepted for publication in the Proceedings of the ICATPP Conference on
Cosmic Rays for Particle and Astroparticle Physics,\\ Villa  Olmo (Como, Italy), 7--8 October, 2010, \\to be published by World Scientific (Singapore).
\end{center}
\vspace{-1.5cm}

\title{PROTON AND ANTIPROTON MODULATION IN THE \\
HELIOSPHERE FOR DIFFERENT SOLAR CONDITIONS \\
AND AMS-02 MEASUREMENTS PREDICTION.}

\author{P. Bobik$^{1}$, M.J. Boschini$^{2,4}$, C. Consolandi$^{2}$,
S. Della Torre$^{2,5}$, M. Gervasi$^{2,3}$, \\
D. Grandi$^{2}$, K. Kudela$^{1}$, S. Pensotti$^{2,3}$ and P.G. Rancoita$^{2}$}

\address{$^{1}$ Institute of Experimental Physics, Kosice (Slovak Republic) \\
$^{2}$Istituto Nazionale di Fisica Nucleare, INFN Milano-Bicocca, Milano (Italy) \\
$^{3}$Department of Physics, University of Milano Bicocca, Milano (Italy) \\
$^{4}$CILEA, Segrate (MI) (Italy) \\
$^{5}$Department of Physics and Maths, University of Insubria, Como (Italy)\\
E-mail: Davide.Grandi@mib.infn.it}

\begin{abstract}
Galactic Cosmic Rays (GCRs) are mainly protons confined in the
galactic magnetic field to form an isotropic flux inside the galaxy.
Before reaching the Earth orbit they enter the Heliosphere and
undergo diffusion, convection, magnetic drift and adiabatic energy
loss. The result is a reduction of particles flux at low energy
(below 10 GeV), called solar modulation. We realized a quasi
time-dependent 2D Stochastic Simulation of Solar Modulation that is
able to reproduce CR spectra once known the Local Interstellar
Spectrum (LIS). We were able to estimate the different behaviors
associated to the polarity dependence of the Heliospheric modulation
for particles as well as for antiparticles. We show a good agreement
with the antiproton/proton ratio measured by AMS-01, Pamela, BESS,
Heat and Caprice and we performed a prediction for the AMS-02
Experiment.
\end{abstract}

\keywords{Heliosphere, Cosmic Rays, Solar Magnetic Field}


\section{Introduction}

The effect of the heliospheric structure on GCRs propagation can
be reproduced by a two dimensional (radius and helio-colatitude)
Stochastic model solving numerically the Parkers's equation
\cite{prot_art}\!. If we do not take into account the effects of
the Earth magnetosphere \cite{mi_jgr}\! modulated fluxes depends
not only on the level of solar activity but also on particles
charge sign and solar magnetic field polarity
\cite{art_midrift}\!. The study of the modulation of $\bar{p}/p$
ratios is particularly important, because it includes explicitly
the combination of charge sign and polarity dependence. The Local
Interstellar Spectra (LIS) used as input of the code, both for
protons and antiprotons, are taken by the Galprop
model\cite{galprop}\!.

\section{Stochastic Model, Parameters and Data Sets}\label{sec:mod}

\subsection{Main parameters of the Model}\label{subsec:par}

The present code simulates the interactions of a GCR entering the
heliosphere which extends up to a fixed distance of about 100 AU from the Sun.

One of the main parameters of the model is the diffusion tensor.
The parallel diffusion coefficient is: $K_{||}$=$k_1\beta
K_{P}(P)(B_{\oplus}/3B)$, where $P$ is the particle rigidity (usually expressed in GV),
$k_1$ is the diffusion parameter discussed in the next section and $K_{P} \approx P$. The perpendicular diffusion coefficient has two
components, radial $K_{\perp r}$, and polar $K_{\perp \theta}$. We
used the relation: $K_{\perp r} = (K_{\perp})_{0} K_{||}$, where
$(K_{\perp})_{0} = 0.05$. We also considered $K_{\perp \theta} =
K_{\perp r}$ in the equatorial region, while we enhanced its value
in the polar regions of the heliosphere \cite{potgieter2000}\!:
$K_{\perp \theta} = 10 K_{\perp r}$.

We used the tilt angle $\alpha$ of the heliospheric current sheet
(HCS) as a parameter for the level of the solar
activity\cite{six}\!: the higher the value of $\alpha$ the lower
the expected GCR flux, for both solar field polarities. Values of
the tilt angle are computed using two different models: the
usual model uses a line-of-sight boundary condition at the
photosphere and includes a significant polar field correction; an alternative
model uses a radial boundary condition at the photosphere,
and requires no polar field correction. As suggested by Ferreira
and Potgieter\cite{ferreira2004}\!, the classical model is used for
periods of increasing solar activity (for example 2007--2012,
AMS-01 data, AMS-02 data), while the new model fits better for
periods of decreasing solar activity (for example 2000--2007, BESS
data).

The three drift components do not depend on external parameters,
except the solar polarity\cite{ste_como}\! (A$>$0 for positive
periods and A$<$0 for negative periods\cite{six}\!). The general
drift expression is locally unlimited for a quasi-isotropic
distribution\cite{lim1,10}\!, therefore we limit all the drift
components below $(\pi /4)v$, which is the spatially averaged
maximum value.

\subsection{Data Sets}\label{subsec:realdata}

We selected GCR proton and antiproton data from 5 different
experiments in order to compare and tune model results:
AMS-01\cite{AMS01}\!, Caprice\cite{caprice}\!,
BESS\cite{bess0,bess1,bess2}\!, HEAT and Pamela \cite{pamela}\!.
The first two experiments took data in a period of positive solar
polarity, BESS in both solar polarities, and the last two in A$<$0
period. The corresponding periods of measurements are: June 1998
(AMS); May 1998 (Caprice); from July 1997 to December 2004 (BESS);
June 2000 (HEAT); and from 2007 to 2008 (Pamela). Solar wind
values for these periods have been obtained from
omniweb\cite{18}\! by 27 daily averages, while tilt angle values
from the Wilcox Solar Laboratory\cite{17}\!. We can estimate
the diffusion coefficient from a long term
study\cite{moskal02,ilya05}\! of neutron monitor measurements and
the \textit{Force Field model}\cite{force}\! (FFM) approach to the treatment of Heliosphere transport of GCRs,
using $\sim$ 40 years of data ending in 2004.
\begin{figure}[hbt!]
\centering
\includegraphics[width=2.2in]{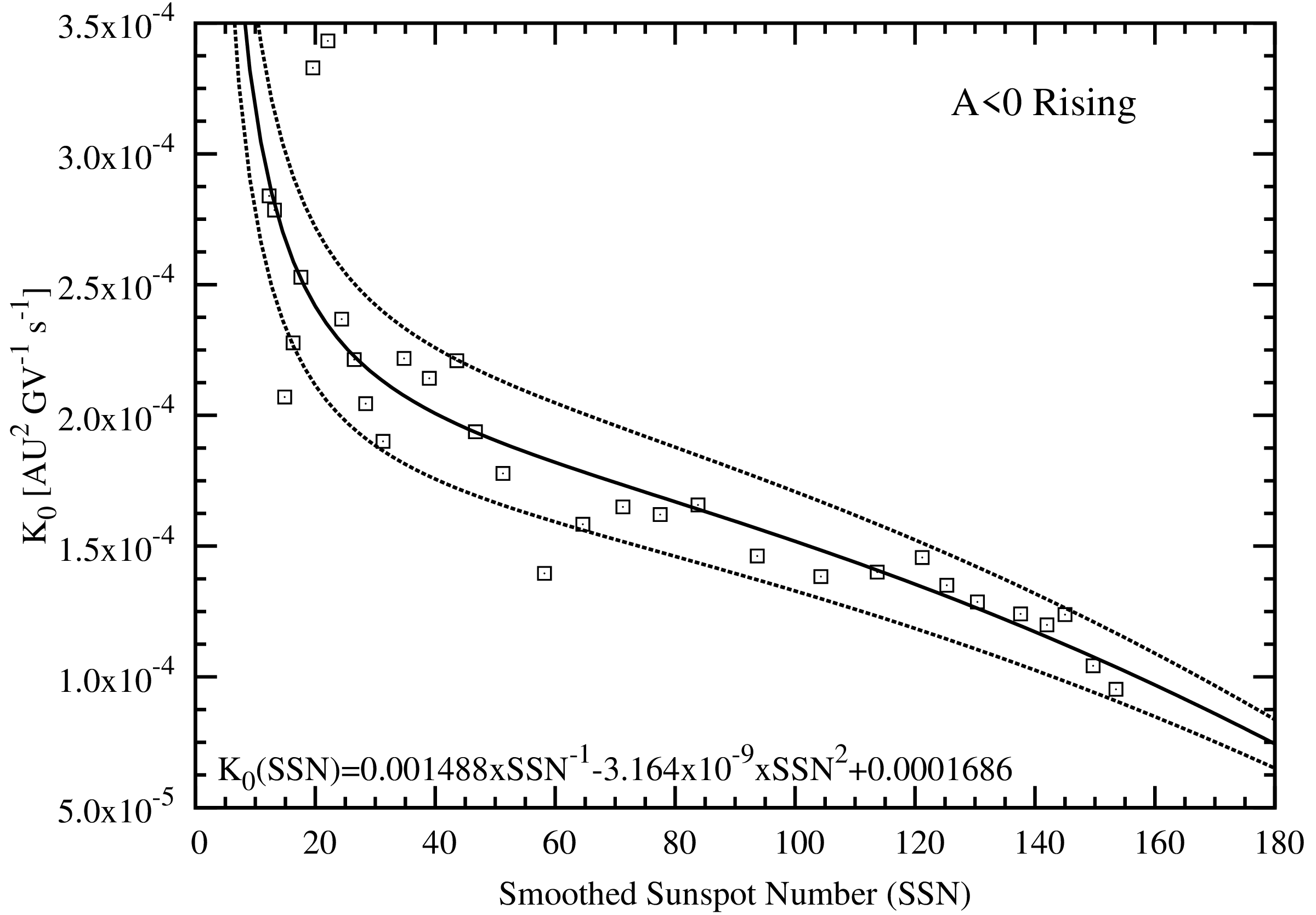}
\includegraphics[width=2.2in]{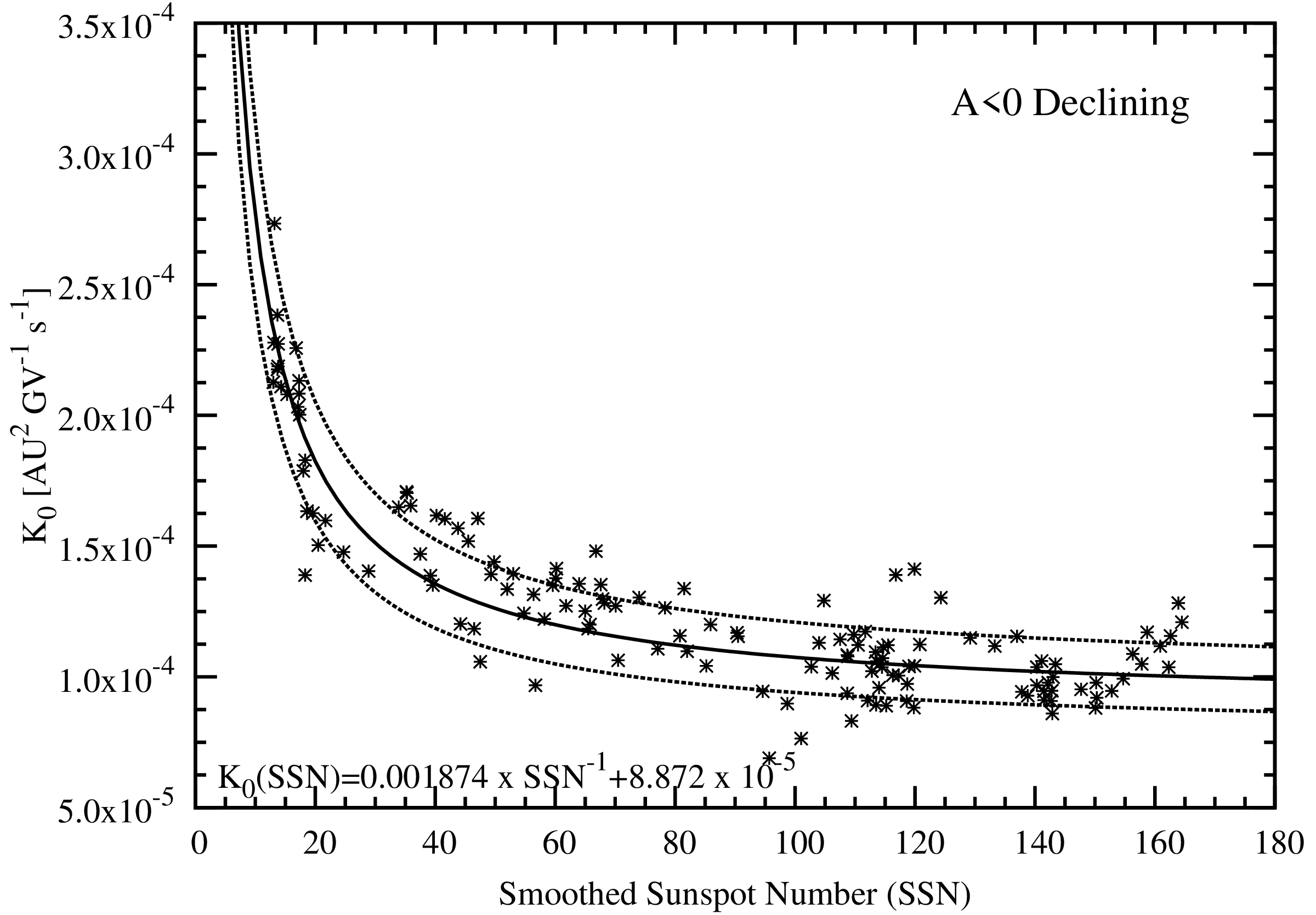}
\includegraphics[width=2.2in]{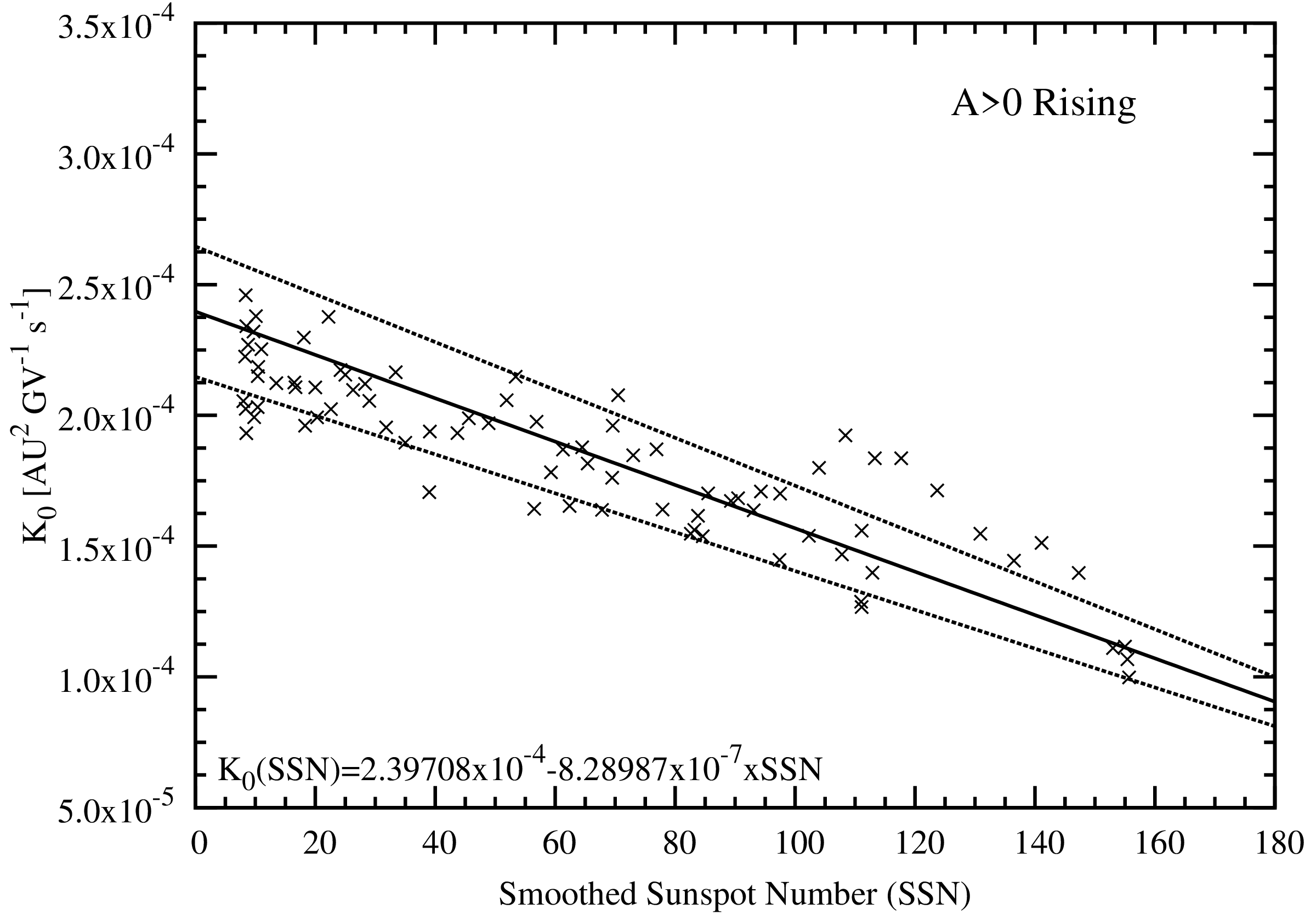}
\includegraphics[width=2.2in]{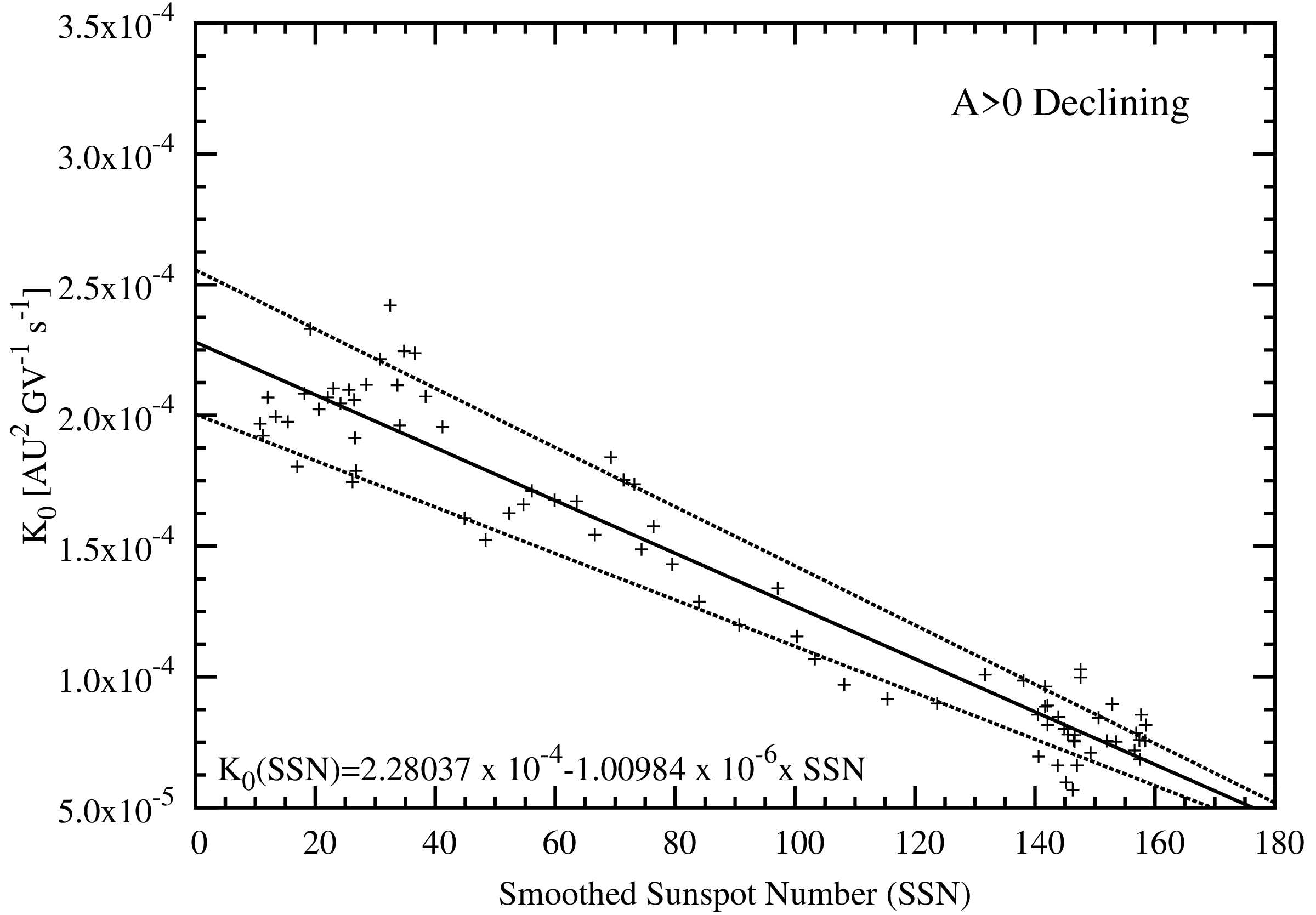}
\caption{Diffusion parameter $K_{0}$ - rising and declining phases for both negative and positive solar magnetic-field polarities - as a function of the SSN value; the continuous lines are obtained from a fit of $K_{0}$ with respect to SSN values.}
\label{fig_sim1}
\end{figure}

\par
In FFM\cite{force}\! [e.g., see also Section 4.1.2.3 of Leroy and Rancoita (2009)], Gleeson and Axford (1968) assumed that i) modulation effects can be expressed with a spherically symmetric modulated number density $U$ of GCRs - the so-called differential density - with kinetic energy between $T$ and $T+dT$, ii) the diffusion coefficient at the time $t$ is given
by a separable function of $r$ (the radial distance from the Sun) and $P$ (the particle rigidity in GV):
\[
\mathcal{K}(r,t) = \beta k_1(r,t) K_{P}(P,t) ,
\]
with $\beta = v/c$ , $v$ the particle velocity, $c$ the speed of light, $K_{P}(P,t) \approx P$ for particle rigidities above $\approx 1\,$GV and iii) the modulation occurs in a steady-state condition, i.e., the relaxation time of the distribution is short with respect to the solar cycle duration so that
\[
\frac{\partial U}{\partial t} = 0.
\]
They derived that the differential intensity at a radial distance $r$ is given by the expression
\begin{equation}\label{force_field_solution}
   J (r,E_{\rm t},t) =  J ({r_{\rm tm}},E_{\rm t}+ \Phi_{\rm p})
   \left[ \frac{ E_{\rm t}^2 -  m_{\rm r}^2 c^4} {( E_{\rm t} + \Phi_{\rm p} )^2-  m_{\rm r}^2 c^4} \right],
\end{equation}
where $J ({r_{\rm tm}},E_{\rm t}+ \Phi_{\rm p})$ is the undisturbed
intensity beyond the solar wind termination located at a radial distance $r_{\rm tm}$ from the Sun; $E_{\rm t}$ is the total energy of the particle with rest mass $m_{\rm r}$ and, finally, $\Phi_{\rm p} $ is the so-called force-field energy loss\cite{force,G_U_1971}\!. When modulation is small\cite{force,G_U_1971}\! - i.e., $\Phi_{\rm p} \ll m_{\rm r}c^2, T$ -, they determined that
\[
\Phi_{\rm p}  = \frac{Z e P}{K_{P}(P,t)}\, \phi_{\rm s}(r,t) \approx Z e \,\phi_{\rm s}(r,t),
\]
where $Z e $ is the particle charge and $\phi_{\rm s}(r,t)$ is the so-called modulation strength (or modulation parameter). Assuming that $v_{\rm w}$ (the solar wind speed) and $k_1$ are almost
constant, $\phi_{\rm s}(r,t)$ - usually expressed in units of GV (or MV) - reduces to
\begin{equation}\label{md_par_red}
     \phi_{\rm s}(r,t) \approx \frac{v_{\rm w}(t) \left( r_{\rm tm} -r  \right)}{3
     k_1(t)} ,
\end{equation}
from which one gets
\begin{equation}\label{md_par_red2}
   k_1(t)   \approx \frac{v_{\rm w}(t) \left( r_{\rm tm} -r  \right)}{3 \phi_{\rm s}(r,t)
    } ,
\end{equation}
i.e., $k_1$ is linearly dependent on $ \left(r_{\rm tm} -r \right)$. In the FFM, the diffusion coefficient $\mathcal{K}(r,t)$ is scalar quantity and, as a consequence, does not account for effects related to the charge sign of the transported particles. $ \phi_{\rm s}(r,t)$ is independent of the species of GCR particles [e.g., see discussion at page 1014 of Gleeson and Axford (1968) or Equation (1) of Usoskin and collaborators (2005)]. The values of the modulation strengths [$\phi_{\rm s}(r_{\rm Earth})$] were monthly determined for the time period\cite{ilya05}\! from 1951 up to 2004 using measurements of neutron monitors (i.e., located at $r_{\rm Earth} = 1\,$AU); while those for the solar wind speeds are available on the web\cite{SW_web}\!. It has to be remarked that $k_1$ depends on the value of the solar wind termination located at a radial distance $r_{\rm tm}$ related, in turn, also to the solar wind speed [e.g., see Chapter 7 of Meyer-Vernet(2007) and Section 4.1.2.2 of Leroy and Rancoita (2009)]. However, because the present simulation code assumes a fixed solar wind termination at 100\,AU to calculate the modulated differential intensities at $r_{\rm Earth}$, one has to derive from the diffusion parameter $k_1$ that ($K_0$) for an effective heliosphere with a radial extension of 100\,AU (see Sect. \ref{subsec:par}). Thus, using Eq.~(\ref{md_par_red2}) one can obtain
\begin{equation}\label{md_par_red1}
    K_0  \approx k_1   \frac{99\,\textrm{AU}}{\left( r_{\rm tm} -r_{\rm Earth}  \right)} = 99\,\textrm{AU}\left[\frac{v_{\rm w} }{3
     \,\phi_{\rm s}(r_{\rm Earth})}\right],
\end{equation}
where 99\,AU is the distance of the Earth from the border of the effective heliosphere as defined in the current simulation code. In Fig. \ref{fig_sim1}, the diffusion parameter $K_{0}$ - obtained from Eq. (\ref{md_par_red1}) - is shown as a function of the corresponding Smooted Sunspot Number (SSN) value.

\begin{figure}[hbt!]
\centering
\includegraphics[width=4in]{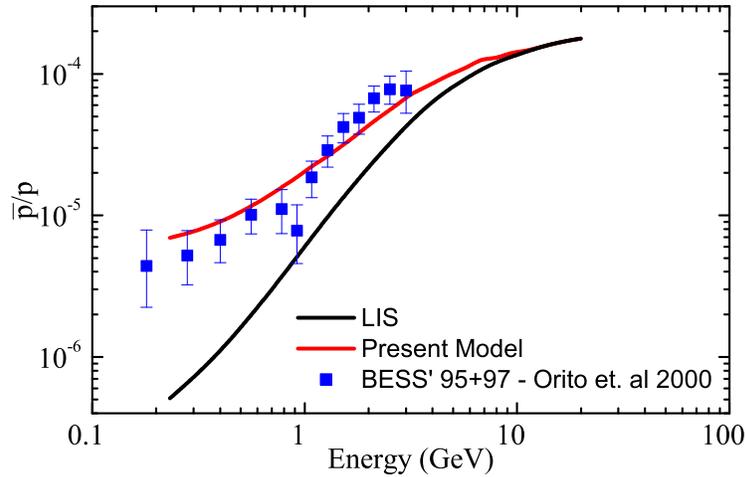}
\caption{Simulated $\bar p / p$ ratio as a function of the kinetic energy in GeV at 1 AU in comparison with
experimental data: BESS (1995-1997).} \label{fig_sim3}
\end{figure}

\begin{figure}[hbt!]
\centering
\includegraphics[width=4in]{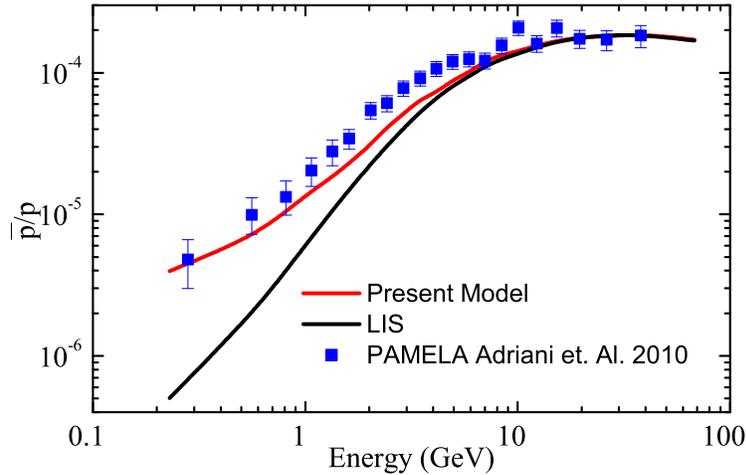}
\caption{Simulated $\bar p / p$ ratio as a function of the kinetic energy in GeV at 1 AU in comparison with
experimental data: Pamela (2007-2008).} \label{fig_sim2}
\end{figure}
\par
The $K_{0}$ data had to be subdivided in four sets, i.e., rising and declining phases for both negative and positive solar magnetic-field polarities. For each set, the data i) could be fitted with a relationship - indicated in in Fi. \ref{fig_sim1} - between $K_{0}$ and Smoothed Sunspot Numbers SSN\cite{ssn}\! values and ii) exhibited a Gaussian distribution of difference of $K_{0}$ values from the corresponding fitted values. The RMSs of the Gaussian distributions were found to be $\approx 0.1339, 0.1254, 0.1040, 0.1213 $ for the phases rising with $A<0$, declining with $A<0$, rising with $A>0$, declining with $A>0$, respectively. In this way - i.e., from the relationships found -, we can use the estimated SSN values
to obtain the diffusion parameter $K_{0}$ at times beyond 2004. In practice, this procedure allows one to extend
the $\approx 40$ years period by exploiting the linear relationship between the fitted $K_0$ values and the SSN values (one of
the main parameters related to the solar activity).
Thus,
we introduced in our code a Gaussian random variation of $K_{0}$
with RMSs corresponding to those found for each subset of data. Results of the simulation with and without the
Gaussian variation are consistent inside the uncertainties of the code \cite{prot_art}\!.

\subsection{Dynamic Parameters}\label{subsec:aver}

Our code simulates the interactions of a GCR entering the
heliosphere from its outer limit, the helio-pause, located - as already mentioned -
approximately at 100 AU\cite{pause}\!, and moving inwards to the
Earth located at 1 AU.
We evaluated the time $t_{sw}$ needed to SW to expand from the
outer corona up to the helio-pause. Considering an average speed
of 400 km/s it takes nearly 14 months. While the time interval
$\tau_{ev}$ of the stochastic evolution of a quasi particle inside
the heliosphere from 100 AU down to 1 AU is $\sim$ 1 month at 200
MeV and few days at 10 GeV. This scenario, where $\tau_{ev} <
t_{sw}$ and $t_{sw} >>$ 1 month, indicates that we can not use
fixed parameters (monthly averages) to describe the conditions of
heliosphere in the modulation process. In fact at 100 AU, where
particles are injected, the conditions of the solar activity are
similar to the conditions present at the Earth roughly 14 months
before. Therefore we consider $\tau_{ev}$ negligible with respect
to $t_{sw}$ and divide the heliosphere in 14 regions, as a
function of the radius. For each region we
evaluated\cite{ecrs08}\! $K_{0}$, $\alpha$ and $V_{sw}$, in
relation to the expansion velocity, in a dynamic way. In the
future the time spent by a GCR particle inside the heliosphere, as
a function of the stochastic path and of the particle energy, will
be also taken  into account.

\begin{figure}[hbt!]
\centering
\includegraphics[width=3.5in]{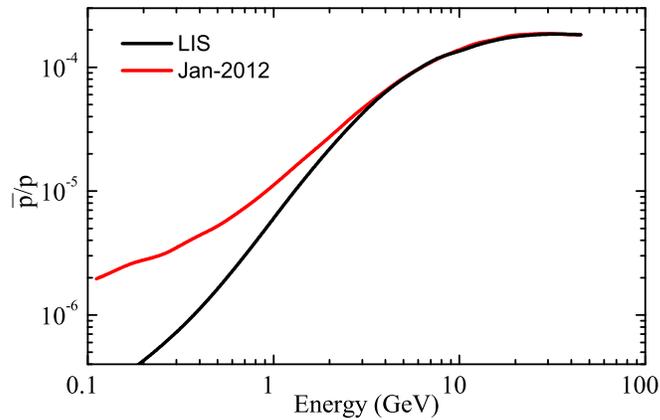}
\caption{Prediction of modulated $\bar p / p$ ratio as a function of the kinetic energy in GeV at 1 AU for
AMS-02 experiment (January 2011).} \label{fig_sim4}
\end{figure}

\subsection{Antiproton/Proton: Comparison with Data and Prediction for AMS-02}\label{sec:ams02}



We performed the simulations using dynamic values of $K_{0}$,
$\alpha$ and $V_{sw}$. Results are shown in \mbox{Fig.
\ref{fig_sim3} and \ref{fig_sim2}}. Simulated fluxes with dynamic
values show a very good agreement with measured data, within the
quoted error bars. This happens both in periods with A$>$0, in
comparison with BESS, and in periods with A$<$0, in comparison with
Pamela. This means that our dynamic description of the Heliosphere
improves the understanding of the complex processes occurring inside
the Solar Cavity.

The periodic behavior of the heliosphere allows us to predict,
with a certain level of precision, the parameters needed for a
simulation related to a time in the near future. In order to get
these data we considered the prediction of SSN from IPS
(Ionospheric Prediction Service) of the Australian Bureau of
Meteorology\cite{IPS}\!.

We concentrate our simulations on the AMS-02
mission\cite{zuccon,casaus}\! that will be installed on the ISS in
February 2011, and, in particular, at a time approaching the solar
maximum: January 2012. We show in Fig. \ref{fig_sim4} the
predictions of GCR modulation for the antiproton/proton ratio.

\section{Conclusions}\label{sec:conc}

We built a 2D stochastic Monte Carlo code for particles
propagation across the heliosphere. Present model takes into
account drift effects and shows quantitatively a good agreement
with measured values, both for positive and negative periods and
for different particles and charge sign. This is relevant because
particles with opposite charge sign undergo a different solar
modulation\cite{art_midrift}\!. We compared our simulations with
antiproton/proton ratios measured by BESS and PAMELA. We used {\it
dynamic} parameters values ($K_{0}$, $\alpha$ and $V_{sw}$) for
the related periods, in order to reproduce the propagation of
incoming GCR through magnetic disturbances carried by the outgoing
solar wind. The {\it dynamic} description of the heliosphere and
the forward approach seem to reproduce better the real physical
propagation of GCR in the solar cavity. In order to have a more
sophysticated model we need to introduce a dependence on the
particle time spent in the heliosphere and a larger statistics of
measured data during negative solar field periods, as AMS-02 will
provide in the next years. Recent measurements\cite{pamela}\! have
pointed out the needs to reach a high level of accuracy in the
modulation of the fluxes, in relation to the charge sign of the
particles and the solar field polarity\cite{mi_ap}\!. This aspect
will be even more crucial in the next generation of
experiments\cite{zuccon,casaus}\!.

\end{document}